\documentstyle[epsf,aps]{revtex}
\draft
\begin{document}
\title{Heisenberg's Uncertainty Relations and Quantum Optics}
\author{G.S. Agarwal}
\address{Physical Research Laboratory,\\ Navrangpura, Ahmedabad-380 009, INDIA}
\date{\today}
\maketitle
\begin{abstract}
We present a brief review of the impact of the Heisenberg uncertainty relations on
quantum optics. In particular we demonstrate how almost all coherent and
nonclassical states of quantum optics can be derived from uncertainty relations.
\end{abstract}
\newpage \noindent
{\bf Introduction: Heisenberg Uncertainty Relations:-} Heisenberg's uncertainty relations had 
tremendous impact in the field of quantum optics particularly in the context 
of the construction of coherent and other classes of states for different 
physical systems and in the reconstruction of quantum states. We present very 
general arguments based on the equality sign in the Heisenberg uncertainty 
relations to demonstrate a very large class of coherent and nonclassical states 
for a wide variety of quantum systems such as single mode and two mode radiation
fields, quantized motion of trapped ions, collection of spins, and two level
atoms. The resulting states for spin systems are especially interesting as these
have varied applications starting from the dynamics of a collection of two level
atoms to two component Bose condensates. Additionally,  spins in such states 
have strong entanglement.\\
Consider a  quantum mechanical system with two physical observables represented
by hermitain operators  $A$ and $B$  satisfying the commutation relation
\begin{equation}
[A, B] =~ iC. 
\end{equation}
One then makes use of the Schwarz inequality
\begin{equation}
\langle(\Delta A)^2\rangle \langle(\Delta B)^2\rangle\geq |\langle\Delta A
\Delta B \rangle|^2,
\end{equation}
and Eq.(1) to obtain the uncertainty relation
\begin{equation}
\langle(\Delta A)^2\rangle \langle(\Delta B)^2\rangle =
|\langle\frac{1}{2}[\Delta A,~\Delta B]\rangle |^2 + |\langle\frac{1}{2}
\{\Delta A, \Delta B \} \rangle |^2 \geq \frac{1}{4}|\langle C\rangle |^2.
\end{equation}
One finds that the equality sign holds only if the state of the system satisfies
the eigenvalue equation,
\begin{equation}
(\Delta A)|{\psi}\rangle = -i \lambda (\Delta B) | \psi\rangle,
\end{equation}
where $\lambda$ is a complex number. We will refer to such states as minimum
uncertainty states. Using  Eq.(4) we can derive some very general results
\begin{equation}
\langle(\Delta A)^2\rangle = |\lambda|^2 \langle(\Delta B)^2\rangle,
\end{equation}
\begin{equation}
\langle\{{\Delta A,\Delta B}\}\rangle=\frac{\rm{Im}\lambda}{\rm{Re}\lambda}
\langle C \rangle.
\end{equation}
Thus for $|\lambda|=1$,~ the observables have equal variance and there is no
correlation between $A$ and $B$ if $\rm{Im}\lambda = 0.$ In what follows we show how
the different coherent and nonclassical states introduced
\cite{{puri},{john},{perelomov},{glauber63},{yuen},{caves},{agarwal},{arecchi},{saran}} in quantum
optics follow from the eigenvalue equation (4). The equation (4) also enables 
us to define in a natural way the squeezed states associated with bosons 
but also for spin systems and other systems such as those described by SU(1,1)
algebra.\\
{\bf System of Bosons:-}
Let us first consider Heisenberg algebra for the position and momentum operators
$x$ and $p$. Then equation (4) leads to a wavefunction $\psi(x)$ which is Gaussian.
On setting $\lambda=1$, we recover coherent states as introduced by
Glauber,\cite{glauber63}
whereas for $\lambda\neq 1$, we recover the squeezed coherent states as
introduced by Yuen and others \cite{perelomov,yuen,caves}.
Note further that the eigenvalue equation (4) in terms of
the harmonic oscillator annihilation and creation operators $a$ and
$a^{\dagger} ([a,a^{\dagger}]=1)$ can be written as 
\begin{eqnarray}
A = {\sqrt\frac{\hbar}{2}}(a+a^{\dagger}),~~ B
&=&{\sqrt\frac{\hbar}{2}}(a-a^{\dagger})/i, \nonumber\\ 
(\mu a + \nu a^{\dagger})|\psi\rangle &\equiv&\alpha | \psi \rangle,\\
\mu=\frac{1}{2}\left(\sqrt{\lambda} +\frac{1}{\sqrt{\lambda}}\right), 
\nu&=&\frac{1}{2}\left(\sqrt{\lambda}-\frac{1}{\sqrt{\lambda}}\right),
\mu^2-\nu^2 = 1.\nonumber
\end{eqnarray}
Thus squeezed coherent state is an eigenstate of the operator obtained by using
Bogoliubov transformed annihilation operator. The states as defined by (7) for
both $\nu=0$ and $\nu\neq 0$ have been extensively used in quantum optics in the
study of radiation fields and the quantized center of mass motion of trapped ions.

In many situations in quantum optics, for example, in dealing with the spontaneous
noise in amplifiers, the state of the field is no longer pure and hence one has
to intruduce a mixed state. In such situations it turns out especially useful to
work with Gaussian Wigner functions of the form \cite{girish}
\begin{eqnarray}
w(q,p)&=&\frac{1}{\sqrt{(2\pi)^2(\alpha\beta-\gamma^2)}} 
\exp\left[-\frac{\alpha q^2+\beta p^2-2\gamma
qp}{2(\alpha\beta-\gamma^2)}\right],\nonumber \\
\alpha &=& \langle p^2 \rangle,~ \beta = \langle q^2 \rangle,~\gamma = 
\frac{1}{2} \langle qp+pq \rangle. 
\end{eqnarray}
Here we assume that $\langle p \rangle = \langle q\rangle=0$. An important feature of (8) is its close
connection to the uncertainty relations. This is because every well defined Gaussian is
{\it not} a bonafide Wigner function. The parameters $\alpha$,~$\beta$
and $\gamma$ must satisfy the stronger form of the uncertainty inequality
\begin{equation}
\sigma = [(\alpha\beta-\gamma^2)^\frac{1}{2} - \frac{1}{2}]\geq 0;
\end{equation}
Note that for a field in thermal equilibrium the correlation parameter
$\gamma=0$, and in addition the uncertainties in $q$ and $p$ are equal 
(setting $m=\omega =1)$. Further one finds a very interesting result for
the entropy associated with the mixed state (8) 
\begin{equation}
S= K_B [(\sigma +1)\rm{In}(\sigma +1)-\sigma\rm{ln}\sigma]
\end{equation}
The entropy thus depends in an important way on the correlation parameter $\gamma$.
The state (8) in fact represents what is now called squeezed thermal state \cite
{adam}, the study of which became very popular in the mid eighties. Furthermore such a state
represents squeezed vacuum for $\sigma=0$. The photon number
distributions corresponding to (8) exhibited a number of interesting features
\cite{adam}.
We further note that states like (8) are
beginning to be used in quantum information theory,  where results like (10)
have been rediscovered for the purpose of calculation of the quantum capacity of
a channel \cite{holevo}.\\
${\bf{\underline{SU(1,1)}} Algebra}$ - {\it Entangled States for Photons, Trapped Ions
etc.}\\ Consider a set of operators $K_i$ satisfying the algebra
\begin{equation}
[K_+, K_-]\equiv -2K_o,~ [K_o,~ K_{\pm}]= {\pm} K_{\pm}.
\end{equation}
This algebra in its various realizations is extremely useful in quantum optics.
In its single mode realization $K_- =\frac{1}{2} a^2,~K_o = a^{\dagger}a+\frac{1}{2}$, the
eigenvalue equation (4) for $A=K_x, B=K_y$ and $\lambda = -1$ leads to even and odd states, 
$|\alpha\rangle\pm|-\alpha\rangle$ with $|\alpha\rangle$ standing
for a coherent state. Such states have been extensively studied for their
non-classical properties. These have been experimentally realized. In fact the
eigenvalue equation (4) for the algebra (11)
\begin{equation}
(\mu K_- + \nu K_+)|\psi\rangle = \lambda|\psi \rangle
\end{equation}
leads to a very wide variety of nonclassical fields associated with the
radiation fields as well as the quantized motion of the center of mass of
trapped ions. For a two mode realization of the SU(1,1) group $K_{-}  =  ab, K_0
\equiv \frac{1}{2}(a^{\dagger}a + b b^{\dagger})$ we get pair coherent states
\cite{agarwal},
\begin{eqnarray}
a b |\Phi\rangle & = &\xi|\Phi\rangle,~(a^{\dagger}a -b^{\dagger}b)|\Phi\rangle =
q|\Phi\rangle\nonumber \\
|\Phi(\xi, q)\rangle &=&  N(\xi,q)\sum_{n=0}^{\infty}\frac{\xi^n}{\sqrt{n!(n+q)!}} 
| n+q,n\rangle,\nonumber \\
N(\xi,q) &=&\{\sum_{n=o}^\infty \frac{| \xi
|^{2n}} {n!(n+q)!}\}^{-1/2} .
\end{eqnarray}
These pair coherent states have very remarkable nonclassical properties such as
entanglement, sub-Poissonian statistics, quadrature squeezing etc. These states have also been
considered \cite{gilchrist} for the study of EPR-like correlations and violation of Bell
inequalities. In Fig.(1) we show the contour plots for the quadrature
distribution $\Phi(x,y)=\langle x,y |\Phi\rangle$ for $\xi = 3.$ Note that the quadrature distribution is far from a Gaussian
distribution \cite{banerji}. Originally such states were shown to occur in the competition of
two nonlinear optical processes, viz; four-wave mixing and two-photon absorption
\cite{gsa}.
More recently a proposal for producing pair coherent states for ionic motion was
outlined \cite{gou}.\\{\bf SU(2) Algebra - Coherent, Squeezed and Entangled States for a 
Collection of Spins}\\
We devote rest of the paper to the problem of different types of states for a
collection of spin systems \cite{{arecchi},{saran},{singh},{veda}}. This would cover a very wide class of systems in
physics. We give some typical examples:-\\ 
{\bf(a) Quantum Properties of Partially Polarized Light} \\
Consider a two mode radiation field with the two modes representing two orthogonal
polarizations of light. The field is in general, partially polarized. We use Schwinger boson
representation to express two modes in terms of spin operators $\vec{S} : 
S^{\dagger} = a^{\dagger} b, S^z = \frac{1}{2}(a^{\dagger}a-b^{\dagger}b)$. The usual optical elements like beam
splitters, wave plates are described by unitary transformations of the form
$\exp(i\theta\vec{S}.\vec{n})$,  where  $\vec{n}$ is a unit vector. Note further that the traditional
Stokes parameters will be given by the expectation value of $\vec{S}$ in the
state of the radiation field.
Since the state of the field could be a nonclassical state, the Stokes
parameters could exhibit very significant quantum fluctuations and these could
be studied in terms of variances like $ \langle S_{\alpha} S_{\beta}\rangle -
\langle S_{\alpha}\rangle\langle S_{\beta} \rangle$. Clearly Heisenberg uncertainty
relations will be especially relevant in the study of quantum fluctuations in
Stokes parameters.\\
{\bf (b) Collection of Two Level Atoms}\\ Such a system plays a very fundamental
role in quantum optics. This system is equivalent to a collection of spins. It
has been shown that squeezed states of a collection of spins can be produced by
interaction with squeezed light i.e. we have a transfer of squeezing from field
to atoms \cite{saran}. Furthermore the dispersive interaction of a collection of atoms in a 
cavity leads to an effective interaction $S^z S^z$ which can lead to the
generation of cat like states for spin systems.\\
{\bf (c) Quantum Dots}\\ Here electron-hole interactions are especially
relevant and a study of collective effects can be done by introducing spin
operators \cite{luis}.
\begin{equation}
S^+ = \sum^{N}_{n=0} e_n^+ h_n^+ .
\end{equation}
The inter dot coupling can be reduced approximately to an effective interaction
 of the form $S^z S^z$, which as we will see leads to the production of
 entangled states.\\
{\bf (d)Two component Bose Condensates}\\ If we approximate each component
of the condensate by a single mode, then the effective Hamiltonian reduces
to \cite{raghavan}\\
\begin{equation}
H = E_a a^{\dagger}a + E_b b^{\dagger}b + \alpha (a^{\dagger} a)^2 + \beta (b^{\dagger}b)^2
+ \gamma a^{\dagger} a b^{\dagger} b^+ (g a^{\dagger}b + g^{\dagger} ab).
\end{equation}
Using the conservation laws, this Hamiltonian can be written in terms of spin
operators as the sum of a quadratic term $\eta S^z S^z$ and linear terms. The 
quadratic term gives us the possibility of the introduction of squeezed and 
other states for a two component Bose condensate.\\
Other examples where states of a collection of spin systems are especially useful include:
(i) Heisenberg exchange interaction, (ii) a suitably oriented state of an atomic
system corresponding to a given $F$ value \cite{klose}. Such a oriented state 
can be produced by suitable optical pumping methods.\\
We start with the eigenvalue equation (4) for the spin problem written in the
form
\begin{equation}
((\vec{a}.\hat{S}) + i\eta (\vec{b}.\hat{S}))\mid\psi\rangle =
\lambda\mid\psi\rangle, 
\end{equation}
where we choose $\vec{a}$ and $\vec{b}$ to be perpendicular to each other. In
particular we can choose $\vec{a} = (\hat{\theta}\cos\phi -
\hat{\phi}\sin\phi)$, $\vec{b}=\hat{\theta}\sin\phi+\hat{\phi}\cos
\phi,$
where $\hat{\theta}$ and $\hat{\phi}$ are the unit vectors in three dimensional
spherical polar coordinate system. Let $\cup(\theta,\phi)$ be the unitary
operator
\begin{equation}
\cup(\theta,\varphi) = \exp(\xi S^{\dagger} - \xi^*S^-),~\xi =\frac{\theta}{2}
~e^{-i\phi}
\end{equation}
The general solution of (16) for $\eta=\pm 1$ is given by\\
\begin{equation}
\mid\psi, \pm\rangle = \cup(\theta,\varphi)\mid S, \pm S\rangle,
\end{equation}
and for $\eta\neq\pm~1~$ by~\\ 
\begin{equation}
\mid \Psi_{m}\rangle={\cal N}_m\cup(\theta,\phi) e^{\mu{S_z}} e^{-i\frac{\pi}{2}S_y}
\mid S,m \rangle,~\eta =\tanh \mu,
\end{equation}
where ${\cal N}_m$ is a normalization constant. The states defined by (18) are the
atomic coherent states which were studied extensively by Arecchi et al \cite
{arecchi}.  The states defined by (19) are the squeezed states for a system of 
spins. These states were studied extensively by Agarwal and Puri 
\cite{{saran},{rrp94}}. It is important to note that both coherent states and the squeezed 
states for a system of spins follow from the Heisenberg uncertainty relations 
for an arbitrary combination of spin operators
$\vec{a}.\vec{S}$ and $\vec{b}.\vec{S}$ with $\vec{a}.\vec{b}=0$.\\
The squeezed states for the spin system as defined by (19) has some remarkable
properties which are similar to the properties for the squeezed states for the 
radiation field. In particular the state $\mid\psi_o\rangle$ for $\theta=0$ is 
the analog of the squeezed vacuum. We summarize some of the important properties
of the state $\mid\psi_0\rangle :$\\
(i) Strong entanglement between different spins in contrast to the states (18)
which show no spin-spin correlations. In particular the state
$\mid\psi_o\rangle$ for $S=1$ has the form
$\alpha\mid\uparrow\uparrow\rangle + \beta\mid\downarrow\downarrow\rangle$.\\
(ii) For integral values of $S,$ the probability $p{(\ell)}$,  of finding the spins in the collective state $\mid S,l\rangle$ is 
an oscillatory function of $l$ and is zero for odd values of $l$.\\
(iii) The phase distribution associated with the state $\mid\psi_{0}\rangle$
exhibits a kind of bifurcation \cite{rps96} which is similar to that exhibited by squeezed
vacuum associated with the radiation field.\\
(iv) The atomic squeezed states can be produced by transferring squeezing from
the radiation field to the atomic system.\\
(v) An important parameter $\xi,$ which is a measure of how much spectroscopic 
resolution can be achieved by using squeezed states, defined by \cite{rrp94}\\
\begin{equation}
\xi=\sqrt{2S}(\Delta S_x)/\mid\langle S_z\rangle\mid
\end{equation}
can be made much smaller than 1.\\
The atomic squeezed states as indicated above exhibit oscillatory population
distributions \cite{saran}. The origin of these oscillations can be traced back to different
pathways that contribute to the amplitude of oscillations \cite{agar96}. 
The zeros in the population distribution are due to complete destructive 
interference between 
two pathways. This is easily seen by using the relation between atomic system
and an equivalent bosonic system. The atomic states can be written in terms of
an equivalent two mode bosonic system as follows:\\
$S^+ = a^{\dagger} b, S^- = ab^{\dagger}, S^z=\frac{a^{\dagger}a - b^{\dagger}b}{2}, a^{\dagger} a+b^
{\dagger} b=$ conserved,\\ 
\begin{equation}
\mid S,m\rangle\leftrightarrow\mid S+m, S - m\rangle_{p}, 
a^{\dagger} a\mid S+m, S-m\rangle_{p} =(S+m)\mid S+m, S-m\rangle_p.
\end{equation}
The amplitude for finding the atoms in the state $\mid S, m \rangle$ in an
atomic squeezed state is proportional to 
\begin{equation}
\langle S,m\mid\psi_0\rangle\propto_{p}\langle S, S\mid\exp\left\{\frac{\pi}{4}\left(a^{\dagger} b-ab ^+\right)\right\}
\mid S+m, S-m\rangle_p.
\end{equation}
The right hand side of Eq.(22) is easily interpreted in terms of a beam splitter
picture - involving incident photons or two ports of the beam splitter.
It is the probability amplitude of finding the outgoing field in the state $\mid
S, S\rangle_p$ given that the input field was in the state $\mid S+m,
S-m\rangle_p$. The complete destructive interference can be understood by 
examining the various pathways leading to $S$ photon on each of the two output
ports.\\ 
{\bf Superposition of Atomic Coherent States:-}  {\bf Cat like states} - We next 
discuss properties and production of cat like states for a collection of spins
in analogy to similar states for the radiation field. i.e. we consider a 
superposition
of atomic coherent states of the form $\mid\theta_1, \phi_1\rangle +\mid
\theta_2, \phi_2\rangle$. The cat like states for bosons have been produced in
different ways,  for example a coherent field passing through a Kerr medium 
leads to the production of cat like states for certain values of the propagation 
length (time)\\ 
We already know that atomic coherent states can be produced by applying a
coherent drive on a collection of spins in ground state. For producing atomic
cat like states, we therefore search for an analog of Kerr medium for the spin
system \cite{singh}.
Let us consider the dynamics of a collection of atoms in a dispersive cavity
described by
\begin{equation}
H =  \hbar\omega_0 S^z + \hbar\omega_c a^{\dagger} a + \hbar g (S^+ a + a^{\dagger}
S^-).
\end{equation}
We assume the cavity to be of high quality and possibly a small number of
thermal photons. We assume large detuning: $\omega_0 - \omega_c\equiv 
\delta_c \gg g.$ Under these conditions the cavity modes can be adiabatically
eliminated and one obtains an effective interaction h
\begin{equation}
h\equiv\hbar\eta\left[\frac{N}{2}\left(\frac{N}{2}+1\right)-S^{z^{2}}
+(2\bar{n} + 1)S^z\right], \eta = \frac{g^2\delta_c}{k^2+\delta_c^2},
\end{equation}
which is quadratic in $S^z$. Such a quadratic term can be interpreted as arising
from the shift of energy levels by cavity vacuum (Lamb shift). It may be added that
a quadratic interaction for an extended system can lead to 
atomic solitons.\\
We next demonstrate how (24) leads to the formation of atomic
or cat states. Consider the time evolution for a system initially in a 
coherent state $(\bar{n}=0)$
\begin{equation}
|\Psi(t)\rangle\equiv e^{{-iht}/\hbar}|\theta,
\phi\rangle=\sum_{K=0}^N {N \choose K}^{1/2} e^{ik\phi}\sin^{N-K}
\left(\frac{\theta}{2}\right)\cos^K\left(\frac{\theta}{2}\right)
{e^{-[i(N-K)(K+1)\eta t]}}
\mid\frac{N}{2}, \frac{N}{2}- K\rangle.
\end{equation}
In order to simplify (25) we note that the exponent which is quadratic
in $K$ has certain periodicity properties. We can write this exponent as a
Fourier series and then each term in the Fourier series can be shown to result in
a coherent state. Calculations show that 
\begin{equation}
\mid\Psi\rangle = e^{\frac{-i\pi N}{m}} \sum^{m-1}_{q=0} f^{(0)}_{q}\bigg | 
\theta,\phi + \pi\frac{2q-N}{m}\bigg>, t=\frac{\pi}{m\eta}, m = odd,
\end{equation}
\begin{equation}
\mid\Psi\rangle = e^{\frac{-i\pi N}{m}} \sum^{m-1}_{q=0} f^{(0)}_q \bigg|\theta,
\phi + \pi\frac{2q-N+1}{m}\bigg>, t=\frac{\pi}{m\eta}, m = even,
\end{equation}
where the $f's$ are the Fourier coefficients. 
We have thus shown how an effective
hamiltonian which is quadratic in the collective operator $S^{z}$ can produce
superpositions of atomic or spin coherent states. Such states exhibit strong
entanglement among individual spins. Note that the production of such entangled
states is also very relevant to systems
like two component Bose condensates and quantum dots, as such systems, under
reasonable approximations,  can be described by an effective Hamiltonian which is
quadratic in $S^z$. In particular for $m=2,$ we generate
a superposition of two coherent states with the same $\theta$, but different $\phi$,
\begin{equation}
\mid\Psi(t)\rangle = \frac{e^{-iN\pi/2}}{\sqrt{2}} \left[e^{i\pi/4}\bigg|\theta,
\phi - \pi\frac{N-1}{2}\bigg> +e^{-i\pi/4}\bigg| \theta,\phi -\pi\frac{N-3}
{2}\bigg>\right].
\end{equation}
Recently considerable success has been achieved with regard to spin squeezing 
\cite{sorensen} both in the context of Bose Condensates and atoms in cells.
Finally, we mention that a general phase space theory for spin systems has been
developed \cite{dowling}. This theory takes into account the conservation of the angular
momentum $S^2$ and its finite length. The applications of this theory to various
problems like EPR correlations have been considered. Besides it is understood how to
reconstruct \cite{wal98} such phase space distributions and the density matrix 
from a variety of measurements.\\

{\bf Atomic Cat States and Multi Atom Greenberg-Horne-Zeilinger States:}
In this section we show the important connection between the Cat state for the
atomic systems and the multi-atom GHZ states. We will demonstrate how (28) is
equivalent to a GHZ state if we assume an initial state $|\theta,\phi\rangle$
which corresponds to full symmetry between different atoms i.e. by choosing 
$|\theta,\phi\rangle$ as 
\begin{eqnarray}
|\theta,\phi\rangle = e^{i N \phi}\Pi_j(\cos\frac{\theta}{2}| g_j \rangle +
e^{-i\phi}\sin\frac{\theta}{2}|e_j\rangle.
\end{eqnarray}
By choosing $\theta = \frac{\pi}{2}$ and $\phi= -\frac{\pi}{2}$, the state
(28) reduces to
\begin{equation}
|\frac{\pi}{2}, -\frac{\pi}{2}\rangle\equiv e^{-iN\pi/2}
|\Psi\rangle_{GHZ},
\end{equation}
\begin{eqnarray}
|\Psi\rangle_{GHZ} &=& \frac{e^{i\pi/4}}{\sqrt{2}} \{\prod_j \frac{1}{\sqrt{2}}(|g_j\rangle
+(-i)^N |e_j\rangle)\nonumber \\
&-& i\prod_j\frac{1}{\sqrt{2}}(|g_j\rangle - (-i)^N | e_j\rangle)\}.
\end{eqnarray}
Note that (31) is exactly a $GHZ$ state. Thus our discussion of last section
also shows how the $GHZ$ states can be produced by using dispersive interaction
in a cavity. This has also been realized in a recent paper by Zheng
\cite{zheng}.\\
Consider next the detection of states like (31) or (28). For this purpose we can
use a Ramsey set up. In the first Ramsey zone (just before atoms enter the
cavity) the atoms are prepared in the state $|\theta,\phi\rangle$. In the second 
Ramsey zone just after the cavity, the cat state (28) gets projected onto 
$\langle \alpha, \beta|\psi\rangle$, where $\alpha$ and $\beta$ refer to the
parameters of the external field in the second Ramsey zone. After the second
Ramsey zone the probability of detecting all atoms in the ground state would 
yield the entangled nature of the state (28) or (31). In particular for $\theta = \pi/2,
\phi=-\pi/2$ and for $\alpha = \pi/2, N=odd,$ Eq.(28) yields a fringe 
pattern, as a function of $\beta$, that is quite different from the corresponding
results in the absence of the cavity and for the incoherent mixture. Note that
one can produce a variety of interference patterns for the state (31) by choosing a range of 
values for the parameters $\alpha$ and $\beta$ referring to the second Ramsey
zone. \\
Clearly Heisenberg's uncertainty relations had a deep impact on the field of 
quantum optics and many independent developments can be viewed from the point 
of a central theme viz. the uncertainty relations.\\
The author thanks some of his long time collaborators and in particular 
R.R. Puri for participating in many projects.\\

\newpage
\begin{figure}
\hspace*{2.4 cm}
\epsfxsize=300pt
\epsfbox{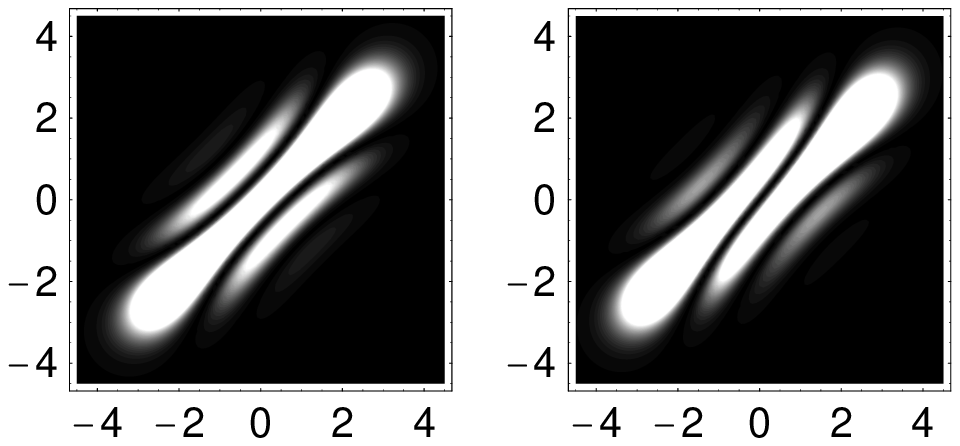}
\vspace*{1 cm}
\caption{The contour plots of the distribution $\mid \Phi (x,y) \mid^2$ for a 
pair coherent state for two different values of the parameter
$q=0(left);=1(right)$.}
\end{figure}
\end{document}